\documentclass[english]{iopart}

  \expandafter\let\csname equation*\endcsname\relax
  \expandafter\let\csname endequation*\endcsname\relax

\usepackage{amssymb,amsmath}
\usepackage[pdftex]{hyperref}
\usepackage{graphicx,caption}
\usepackage{epstopdf}

\usepackage{color}
\usepackage{colortbl}
\usepackage{xcolor} 

\usepackage[integrals]{wasysym}
\usepackage{mathrsfs}
\usepackage{marvosym}
\usepackage{fancybox}
\usepackage{fancybox}
\usepackage{dcolumn}
\usepackage{bm}

\usepackage[T1]{fontenc}   
\usepackage{selinput}      
\SelectInputMappings{
	adieresis={ä},
	germandbls={ß},
	Euro={€}
}

\begin{document}

\title{BEC array in a Malleable Optical Trap formed in a Traveling Wave Cavity}

\author{D. S. Naik, G. Kuyumjyan, D. Pandey\footnote{Now at: Institut f{\"u}r Angewandte Physik der Universit{\"a}t Bonn, Wegelerstraße 8, 53115 Bonn, Germany}, P. Bouyer, A. Bertoldi}
\address{LP2N, Laboratoire Photonique, Num{\'e}rique et Nanosciences, Universit{\'e} Bordeaux--IOGS--CNRS:UMR 5298, rue F. Mitterrand, F--33400 Talence, France.}
\ead{andrea.bertoldi@institutoptique.fr}

\date{\today}

\begin{abstract}
Although quantum degenerate gases of neutral atoms have shown remarkable progress in the study of many body quantum physics, condensed matter physics, precision measurements, and quantum information processing, experimental progress is needed in order to reach their full potential in these fields. More complex spatial geometries as well as novel methods for engineering interesting interactions are needed. Here we demonstrate a novel experimental platform for the realization of quantum degenerate gases with a wide range of tune-ability in the spatial geometries experienced by the atoms and with the possibility of non-trivial long-range interactions both within and between multiple $^{87}$Rb Bose-Einstein condensates (BECs). We explore the use of a large mode-volume bow-tie ring cavity resonant at two wavelengths, $\lambda$=1560 and 780 nm, for the creation of multiple BECs within a Malleable optical trap which also possesses the ability of photon-mediated long-range interactions. By exciting diverse transverse modes at 1560 nm, we can realize many optical trapping geometries which can open the door to spatial quantum state engineering with cavity-coupled BECs. As representative examples we realize a BEC in the fundamental TEM$_{00}$ and a double BEC in the TEM$_{01}$ mode of the cavity. By controlling the power between the fundamental and the higher transverse cavity mode, splitting and merging of cold thermal atomic ensemble is shown as well as the potential of creating more complex trapping geometries such as uniform potentials. Due to the double resonance of the cavity, we can envision a quantum network of BECs coupled via cavity-mediated interactions in non-trivial geometries.

\end{abstract}


\maketitle

\section{Introduction}
In the past fifteen years, many phenomena of long-standing interest in condensed-matter physics have been realized in ultra-cold atomic settings \cite{Bloch_2008}, providing considerable advantages compared to condensed-matter systems: ultra-cold atomic systems are highly controllable, well isolated from environment and governed by thoroughly understood tunable Hamiltonians (e.g., Feshbach tuning of interaction strength \cite{Chin_2010}, optical lattice depth \cite{Morsch_2006}, etc. are governed by quantities easy to alter such as magnetic fields and laser intensities, respectively). The high degree of control and tuneability afforded by optical potentials has made it possible both to explore phenomena in a simpler setting than is typical in condensed matter and to address hitherto experimentally inaccessible questions, such as the dynamics of ordering in systems that are quenched past a quantum critical point \cite{Greiner_2002}, the physics of electrons propagating through static lattices \cite{Jordens_2008}, the construction of novel quantum fluids \cite{Paredes_2004,OHara_2002} and of non-equilibrium phase transitions \cite{Kollar2017}. \\

However, many geometries of great interest in condensed matter physics are difficult either to realize or to load and cool atoms into, e.g. uniform potentials \cite{Gaunt_2013}, Josephson double wells, as well as dynamical time varying potentials such as multi-port atomic beamsplitters \cite{PhysRevLett.98.110402,Schumm2005}. Creating more complex spatial geometries usually involves multiplexing multiple optical beams alongside advanced beam shaping technology, like spatial-light modulators \cite{Gaunt_2013} or two-dimensional spatial modulation schemes with acousto-optic modulators (AOMs) \cite{Meyrath_2005} involving fast switching and a high degree of experimental control. Here we propose a simpler, more robust scheme to engineer high power optical potentials with easily controllable spatial properties which builds on the platform of many-body Cavity QED, relying on linear combinations of transverse spatial modes of a bow-tie cavity to trap atoms in desired potential geometries. In principle, optical potentials of any even powered spatial profile or combinations of even powered spatial profiles can be realized. Moreover, atoms can be adiabatically transported from one trapping geometry to another. We refer to these types of potentials as Malleable optical potentials and we show how they allow the preparation of ultra-cold atoms with high phase-space density in unique trapping geometries.\\

\begin{figure}[h]
\centering
\includegraphics[width=0.60\textwidth]{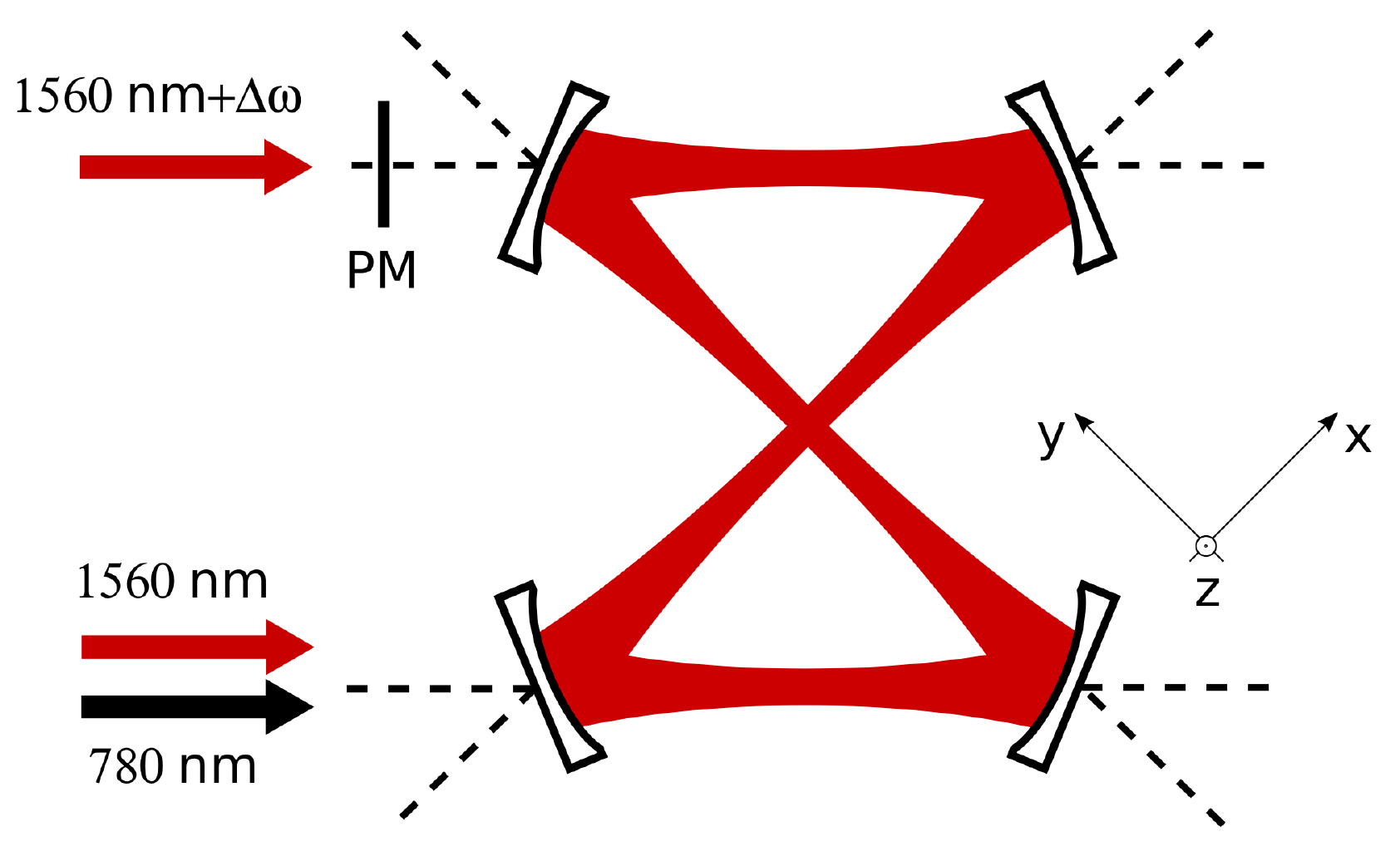}
\caption{Our bow-tie cavity has multiple input/output ports available to independently couple and excite different transverse cavity modes. By locking the main 1560 nm beam to the cavity and using fiber beam splitters and fiber AOMs we can create multiple input channels each of which can be frequency tuned to the relevant spatial mode that we wish to excite via the multiple input/output ports (see Fig. \ref{trans_nondeg} for frequency shift values). An additional spatial phase mask (PM) is used to increase the coupling to higher transverse modes.}
\label{fig:cavita}
\end{figure}

In this article, we experimentally realize Malleable optical potentials for BECs, and also show the potential of creating multiple independent condensates coupled by a common cavity mode, which can be used to engineer controllable interactions between the condensates \cite{Byrnes_2013,Vaidya2018} or their spin components \cite{Landini2018}. We first describe our experimental platform, a dual-frequency traveling wave cavity, focusing on the ability to create an array of multiple BECs using higher transverse spatial modes of the cavity \cite{Kruse2003}, and characterize the single and double BECs that have been created. Then we describe how this system can be used to create arbitrary potential geometries, focusing on two specific realizations: 1) a quantum beam-splitter for BECs and 2) a uniform potential. A more detailed description of our setup is included in the appendix. We will finally discuss various schemes to engineer long-range interactions between the different condensates.

\section{Malleable Optical Potentials created in a Cavity}
The heart of the experiment is a large-volume optical bow-tie cavity, simultaneously resonant at both 1560 nm and 780 nm. The large-volume allows us to create a magneto-optical trap (MOT) and evaporate multiple BECs directly in the center of the cavity, thereby bypassing the need for complex atom transport schemes. Due to the multiple input/output ports afforded by our bow-tie cavity configuration and the non-degeneracy between the TEM modes, we can simultaneously and selectively excite multiple TEM modes of the cavity, allowing the creation of more complex trapping geometries in one-, two-, and three-dimensions. This offers several advantages: 1) good spatial mode quality of the trapping potential as the cavity serves as a spatial filter, 2) high trapping potentials due to the build-up in intra-cavity power, 3) good relative alignment and minimal positional noise of optical modes (reducing heating) due to the cavity structure, and 4) trapping geometries can be easily changed adiabatically, transporting atoms from one geometry to another.

The cavity resonance at 1560 nm allows the production of a deep far-off-resonance optical trap (FORT) for our atoms using a low-power laser source via cavity enhancement of light inside the resonator. The presence of the resonance at 780 nm opens up the possibility to enhance atom-light interactions with the goal of improving the efficiency of non-demolition measurements in measurement--and--correction schemes \cite{Vanderbruggen_2013,Kohlhaas_2015}. In addition, the expected linewidth at 780 nm of our large-volume cavity is of the order of the single-photon recoil shift, opening the door to cavity-assisted cooling \cite{Wolke_2012,Sandner2013} and the creation and manipulation of momentum entangled states \cite{Cola_2009,Bux_2011,Piovella_2012,Kessler2014}.

\subsection{Cavity Mode Structure} \label{cavitystructure}
The electric fields inside our cavity can be naturally decomposed into the orthonormal Hermite-Gaussian modes \cite{Siegman_1986}. For the beam propagating along the $x$ axis (see Fig. \ref{fig:cavita}), and with the origin at the center of the configurations one obtains:

\begin{align}
E_{mn} & = & E_0 \frac{w_0}{w(x)}  H_m \left( \frac{\sqrt{2} \, y}{w(x)} \right) \exp \left( - \frac{y^2}{w(x)^2} \right) H_n \left( \frac{\sqrt{2} \, z}{w(x)} \right) \exp \left( - \frac{z^2}{w(x)^2} \right) \nonumber\\
& & \times \exp \left( -i \left[k_z - (1+n+m) \arctan\left(\frac{z}{z_R}\right) + \frac{k(y^2+z^2)}{2R(z)} \right] \right)
\end{align}

where $H_n(x)$ is the Hermite polynomial with the non-negative integer $n$. The indices $m$ and $n$ determine the shape of the profile in the $y$ and $z$ direction, respectively. The quantities $w$ and $R$ evolve in the direction of propogation ($x$).

For a standard two-mirrors non-degenerate optical cavity, two transverse modes, TEM$_{mn}$ and TEM$_{lk}$, given by $m+n \neq l+k$ are non-degenerate, whereas the modes given by $m+n = l+k$ are degenerate. However, due to the folded nature of our cavity, reflections on the plane of the cavity occur off-axis on the mirrors, leading to an astigmatism between the horizontal and vertical modes, with the result that all modes are non-degenerate. Therefor we can independently excite any transverse mode (or linear combination of transverse modes) by frequency and spatial matching of the input beams to the desired modes (Fig. \ref{trans_nondeg}).

As mentioned in \ref{FORT}, all 1560 nm light originates from one laser. Using fiber beamsplitters, a small portion is taken and amplified to 3 W to form the TEM$_{00}$ injection beam. This laser is locked to the cavity via the Pound-Drever-Hall method \cite{Drever_1983}. The light coming from the other ports of the the fiber beam-splitters may be used to excite higher order transverse modes by frequency shifting them via AOMs with respect to the TEM$_{00}$ beam by the relevant amount, and shaping their phase profile by means of a phase mask to improve the coupling efficiency. These beams can be amplified and injected through the multiple input ports of the cavity. In addition, the AOMs can be used to control the power of each mode. We can thus realize a linear combination of excited modes:
\begin{equation}
\bm{E}_{tot} =  \sum_{m,n} \sqrt{\beta_{mn}} \bm{E}_{mn}
\end{equation}
where $\beta_{ij}$ is given by the power in each mode. Due to the complete transverse non-degeneracy of the cavity and the orthogonality of the different Hermite-Gaussian modes (see \ref{app:orthogonalityModes}), the intensity can be written as 
\begin{eqnarray} \label{eq:intprofile}
I_{tot} &=&  \frac{c n \epsilon_0}{2} \lvert \bm{E}_{tot} \rvert^2 \nonumber\\
        &\approx&  \frac{c n \epsilon_0}{2} \sum_{m,n} \beta_{mn} I_{mn}
\end{eqnarray}

The $\beta_{mn}$ terms can be modulated as fast as the cavity linewidth and the control AOMs permit, allowing us to perform fast quenches of the trapping potentials which can be used to study out-of-equilibrium quenched quantum systems \cite{Eisert_2015} and also allow us to implement atom optics manipulation tools for condensed samples (see Sec. \ref{beamsplitter}).

It should be noted that in principle at the crossing region of the cavity, a single mode in the different arms may interfere with itself to form a standing wave. We eliminate this possibility by injecting the cavity with laser polarization that is in the plane of the cavity, i.e. horizontal. Thus in the crossing region, the two orthogonal arms meet with orthogonal polarizations, thus avoiding interference patterns.

\subsection{Towards a Quantum Network of BECs} \label{Exp_BECs}

The cavity in our setup provides the possibility to create arrays of BECs in a line, a plane or a 3 dimensional geometry due to the feasibility of exciting higher order transverse cavity modes. A particular transverse mode TEM$_{\textrm{m,n}}$ corresponds to $(\textrm{m}+1)^2 \times (\textrm{n}+1)$ trapping regions at the cavity crossing point \cite{Bertoldi_2010}. The availability of the multiple input-output ports results in an additional advantage for the folded configuration of the cavity. The simultaneous excitation of two modes, by injecting appropriately frequency and phase matched beams allows the transfer, splitting and recombination of atoms between two different trapping configurations. First we present the creation of BECs for two specific cases, cavity injected with the TEM$_{00}$ and TEM$_{01}$ modes, which are the single and the double well trapping configuration. Then we will discuss the capabilities of transferring atoms from the TEM$_{00}$, where established techniques of evaporative cooling afford efficient cooling to BEC, to a number of various more interesting optical trapping geometries where evaporative cooling may be less efficient (i.e. see Sec. \ref{UniformPot} on uniform potentials).\\

\begin{figure}[h]
\centering
\includegraphics[scale=0.5]{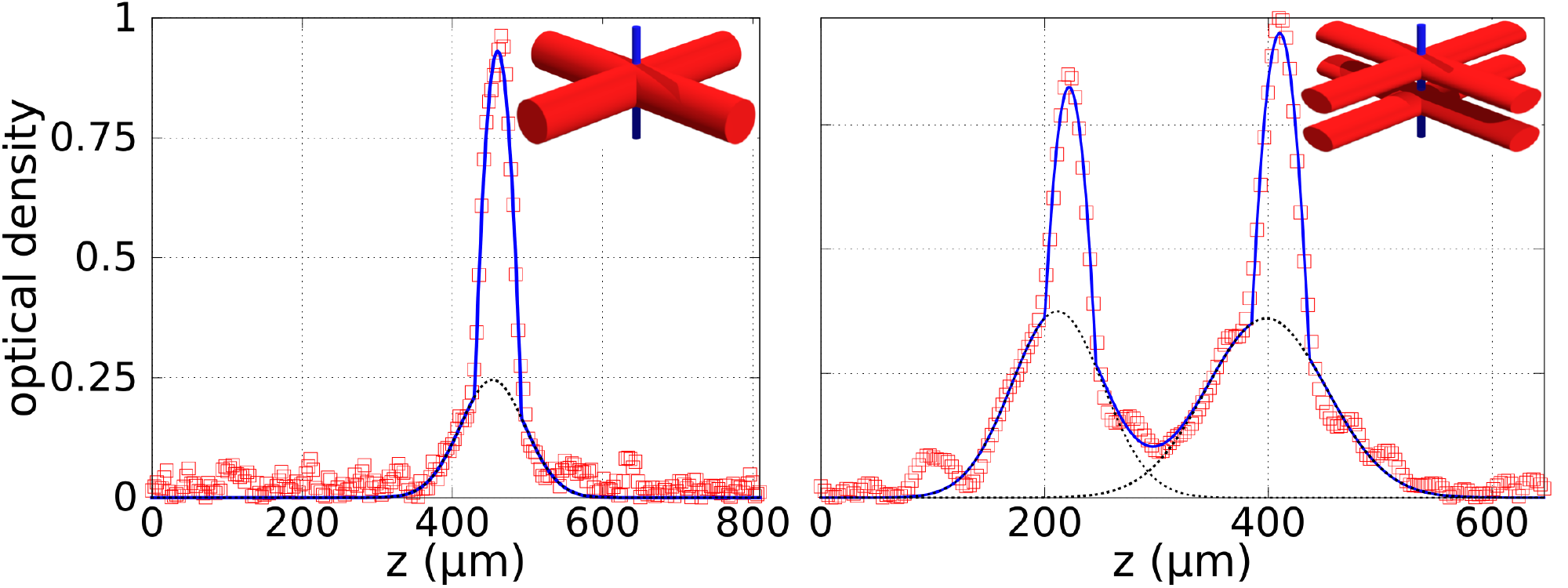}
\caption{BEC in the TEM$_{00}$ (left, condensed fraction 52\%) and the TEM$_{01}$ (right, condensed fraction of 27\% and 29\% for the lower and higher cloud respectively) mode of the cavity. Shown are the horizontal integration of the absorption images taken 8 ms after the atoms are released from the FORT (red squares), and their fit with a single (left) and double (right) Thomas-Fermi model (blue solid line). The thermal fractions resulting from the fit are shown with black dashed lines. The insets show the cavity configuration used in each case, i.e. TEM$_{00}$ (left) and TEM$_{01}$ (right), with the vertical dimple in blue at the center of the configuration.}
\label{BEC00}
\end{figure}

\underline{\textit{BEC in TEM$_{00}$ mode}} An in-depth description of our experimental setup for cooling and trapping atoms at the cavity center using a 2D/3D MOT setup can be found in \ref{exp_setup}. At the end of the loading sequence, about 2$\times$10$^6$ atoms can be found loaded into the far-off-resonance optical trap (FORT) created in the cavity crossing region (see Fig. \ref{BEC00} Left Inset). After thermalization, the sample temperature is 200 $\mu$K, leading to a temperature to trap depth ratio of k$_B$T/U$_0$ $\sim$ 1/6. Due to strong one-body losses lifetime of trapped atoms is limited to 6.6 s.

The transition temperature of the BEC is 180 nK which is achieved by forced evaporative cooling of the trapped thermal ensemble in the intra-cavity FORT. During the entire MOT and evaporation process, a tightly focused 1560 nm dimple beam with waist $w_0 \sim 23$ $\mu$m, radial and axial frequencies $\omega_{\mathrm{r}}\sim 2 \pi \times 3$ kHz and $\omega_{\mathrm{\perp}}\sim 2 \pi  \times 50$ Hz, is used in the vertical direction, intersecting the center of the crossed optical dipole trap but focused slightly above. The stagnation of forced evaporation near the end of the ramp, as a result of the reduction in trap frequencies during the lowering of the optical trap, requires the dimple to enhance the scattering cross-section at the end of the evaporation process by increasing the trapping frequencies along the plane of the cavity (the vertical axis is weakly affected by the dimple) \cite{Dalibard_2011}. The dimple has an added advantage of allowing us to compensate for the effect of the linear slope of gravity (see below). After thermalization of the atoms in the FORT, the intra-cavity trapping power is adiabatically weakened following the empirically optimized form of $U_c (1+t/{\tau})^{-\beta}$ with $\tau$=180 ms, $1.4 < {\beta} < 1.8$ for a duration of around 3.2 s to produce a BEC of 3$\times 10^{4}$ atoms in TEM$_{00}$. The phase space density at this stage is $> 2.7$ with the peak atomic density $n_0 \sim 6 \times 10^{\mathrm{13}}$ atoms/cm$^3$.

\begin{figure}[h]
\centering
\includegraphics[width=0.5\paperwidth]{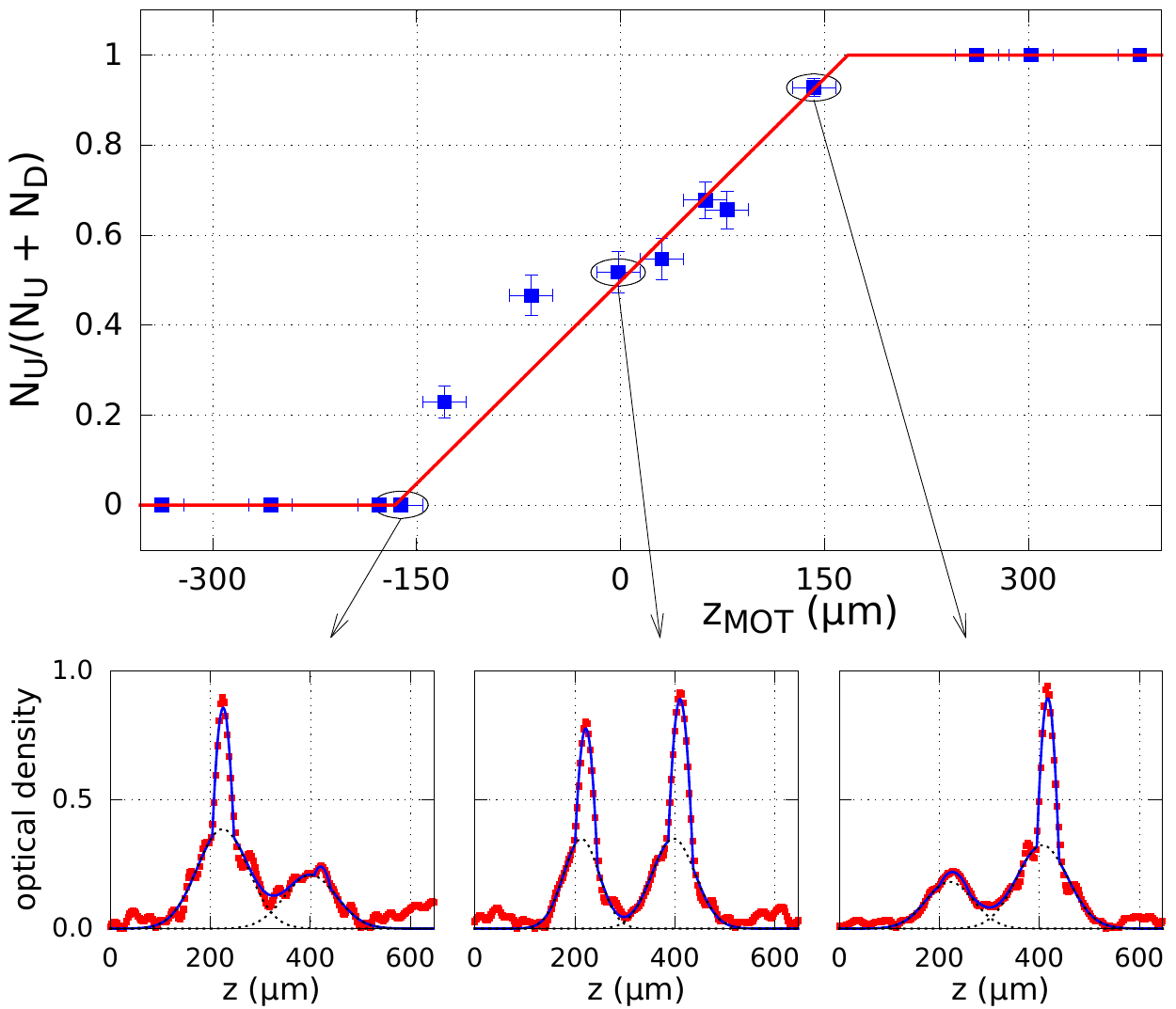}
\caption{Number of condensed atom in the upper potential well (N$_U$), normalized with the total number of condensed atoms in the two wells (N$_U$+N$_L$, where N$_L$ is the number of condensed atoms in the lower well), as a function of the vertical position where the quadrupole field of the MOT vanishes. Each point results as an average over 3 experimental samples. The MOT position is varied by changing the current in the coils used to compensate the vertical bias magnetic field, controlling in this way the number of atoms loaded in each well. The fit (red line) consists of a linear function with negative slope, constrained at 0 and 1 outside the $\left[ 0:1 \right]$ interval. The insets show typical magnetic field configurations resulting in a BEC mainly in the lower well (left inset), equally distributed (center inset) and mainly in the upper well (right inset). The horizontally integrated optical densities (red circles) are integrated with the sum of two independent Thomas-Fermi distributions (blue solid lines), whose thermal components are shown in black dashed lines.}
 \label{Bz1}
\end{figure}

Fig. \ref{BEC00} (left panel) shows the vertical profile of the optical density integrated along the horizontal direction (red squares); the data are fitted with a Thomas-Fermi model (blue line) and show 52\% of the atoms in the condensed fraction. The absorption image is taken 8 ms after releasing the atoms from the trap (with trapping frequency $\omega \sim 2 \pi \times 200$ Hz). A sharp parabolic feature on the top of the broad Gaussian profile exhibits the degenerate nature of the ultra-cold gas.\\

\underline{\textit{Double BEC in TEM$_{01}$ mode}} If the entire procedure is repeated but with the TEM$_{01}$ mode of the cavity injected, the crossing region will consist of the intersection of the TEM$_{01}$ mode with itself, leading to two trapping regions vertically separated (see Fig. \ref{BEC00} Right Inset). In this manner two BECs each with 2$\times 10^{4}$ were obtained. The intra-cavity power will now be shared between two crossing points, whereas the single dimple beam is aligned to cross both trapping regions; this configuration leads to the same trap frequencies in the plane of the cavity with respect to the case of the TEM$_{00}$ mode (see Fig.\ref{BEC00}).

Fig. \ref{BEC00} (right panel) shows the double BEC obtained when the cavity is injected with the TEM$_{01}$ mode. This creates two trapping regions along the vertical z-axis on the two intensity maxima of the Hermite-Gauss function, with a separation of around 140 $\mu$m. Fig. \ref{BEC00} (right panel) shows the BECs in this double well configuration released from the trap with average frequencies $\omega \sim 2 \pi \times 70$ Hz, after a time of flight of 8 ms; the condensed fractions in the lower and upper clouds are 27\% and 29\% respectively. The BECs obtained both in the TEM$_{00}$ mode and in the TEM$_{01}$ one present a spatial displacement between the center of the condensed and the thermal fraction, which results from the anharmonicity of the trapping potential \cite{Klinner2010,Bertoldi2010a}.

The population ratio of the BECs can be controlled by applying a vertical magnetic field bias during the MOT loading phase, leading to a different loading efficiency of the thermal atoms into each individual well (see Fig. \ref{Bz1}). For a certain range of bias field one can have BECs with an equal number of condensed atoms in both the traps.

As the different trapping wells are loaded from the MOT independently, the resultant BECs are independent, however they exist within the same cavity, and could then be coherently coupled via a common cavity mode. The cavity resonance at 780 nm could be exploited to perform coherent operations between the BECs.\\

\begin{figure}[!h]
\centering
\includegraphics[width=0.65\textwidth]{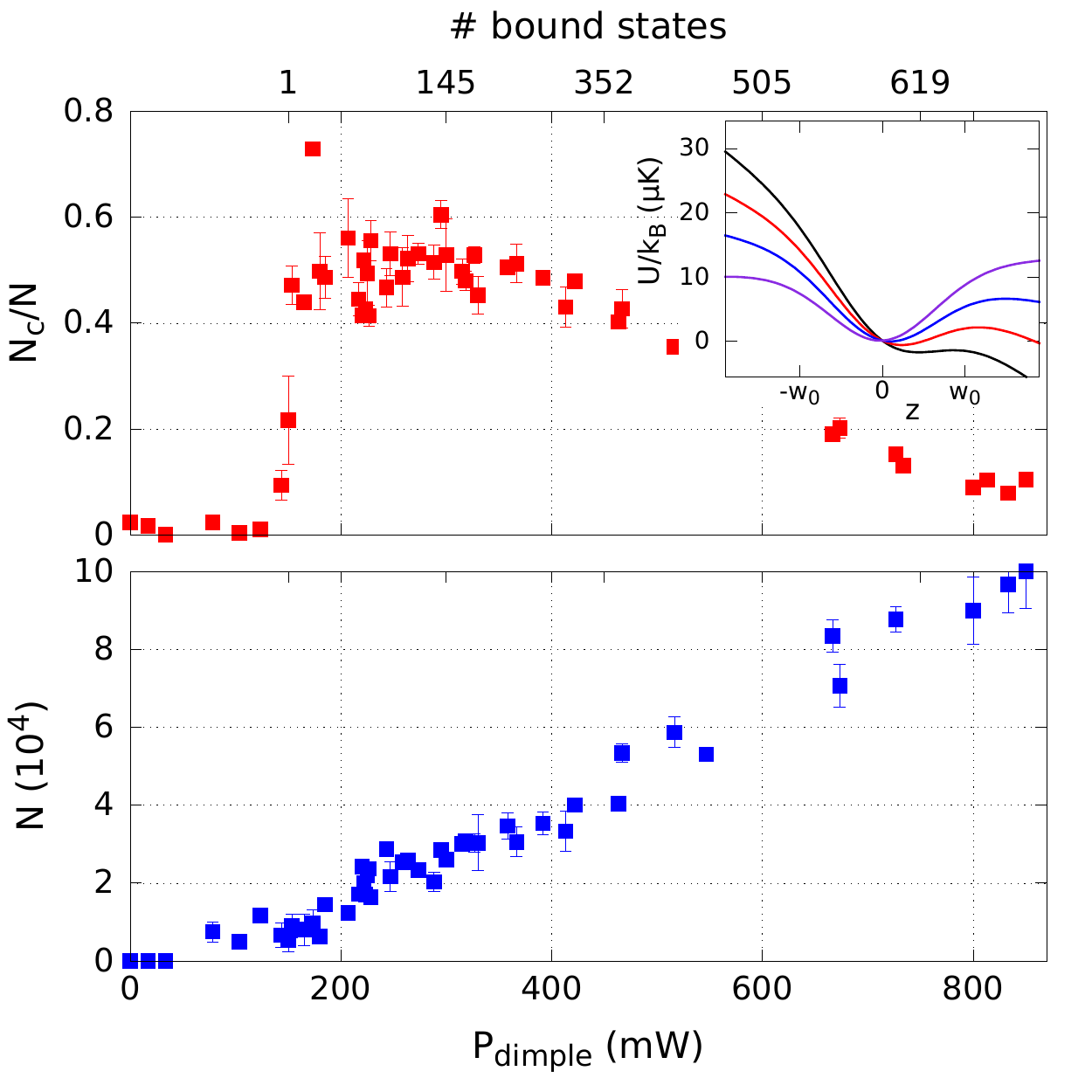}
\caption{Condensate fraction N$_c$/N and total atom number N as a function of dimple beam power P$_{\mathrm{dimple}}$ for a given evaporation ramp. Each experimental point is the average over 3 repetitions. The dimple is vertically offset to where varying the dimple power will not effect the curvature of the resultant trap (and therefore the number of bound states of the trap) but the vertical potential on the wings of the dimple can counteract the linear effect of gravity, thus creating a combined potential that is deeper (and thus has more bound states as plotted on the top vertical axis of the graph). With a dimple power of 150 mW, our combined trapping potential supports one bound state, and therefore we see the formation of a large condensate, but with low atom numbers. Increasing the dimple power reduces the linear effect of gravity, as shown in the inset for different dimple power (P$_{\mathrm{dimple}}$ = 0 (black), P$_{\mathrm{dimple}}$ = 1.32 W (red), P$_{\mathrm{dimple}}$ = 2.64 W (blue) and P$_{\mathrm{dimple}}$ = 4 W (violet); $w_0$ indicates the vertical waist of the cavity). In this way we can increase the number of available bound states for the atoms, thus reducing the condensate fraction but also increasing the number of total atoms as our trap depth is larger. Note, the axis designating the number of bound states is not linear as the dependence of the number of bound states on the dimple power is not linear.}
 \label{dimple}
\end{figure}

\underline{\textit{Role of the Dimple}} The dimple beam plays a vital role to obtain the BEC both in the TEM$_{00}$ mode and in the TEM$_{01}$ one, by increasing the trapping frequency and the collision rate between the atoms during the end of the evaporation process in the plane of the cavity (vertical frequencies are weakly effected by the dimple). However, the dimple has an additional, subtle but important effect again at the end of the evaporation stage. The linear gravitational potential effectively decreases the trap depth in the vertical direction. We displace the focus of the dimple vertically from the cavity center, to where the atoms feel a linear potential due to the intensity reduction away from the focus. This linear potential arising from the displaced dimple beam serves to reduce the linear potential due gravity. By controlling the power of the dimple, we can vary the effective linear potential seen by the atoms, even canceling the effect of gravity \cite{Ricci_2007} with enough power (see Fig. \ref{dimple} (inset)). By partially canceling this linear potential via the dimple beam, we can increase the trap depth and thereby change the number of quantum levels inside our trap. This is evidenced by the condensate fraction at the end of the evaporation sequence with respect to dimple power (see Fig. \ref{dimple}). With a dimple power of 150 mW, we arrive at a trapping condition which allows at most one quantum level in the trap, and therefore a large condensate fraction but low overall atom numbers due to the small trap depth. By increasing the dimple power, we can increase the number of quantum levels for the atoms to occupy, increasing the number of atoms due to the larger trap depth but decreasing the condensate fraction as these atoms have more states to distribute themselves into.

\begin{figure}[h]
\centering
\includegraphics[width=0.65\textwidth]{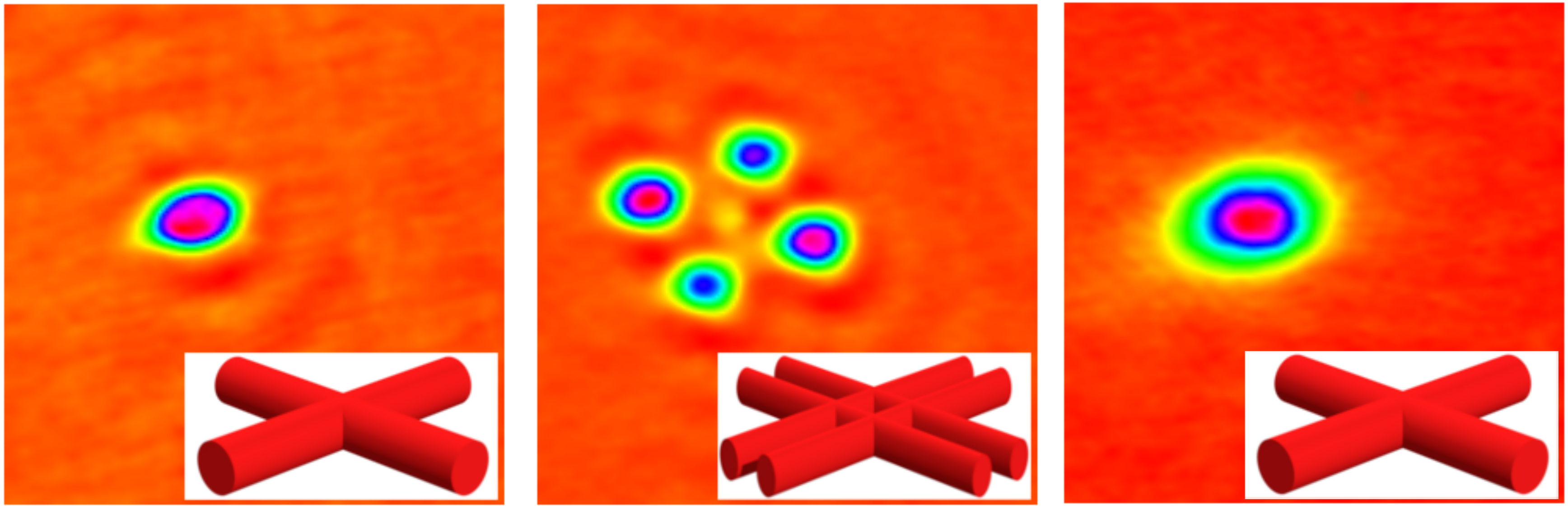}
\caption{Splitting of a single cold thermal ensemble trapped in the TEM$_{00}$ mode to four cold ensembles in the TEM$_{10}$ mode of the cavity (left to middle panels) and the reverse process of recombination (middle to right panels) by controlling the relative intensity between the two modes. The ensemble temperature in the TEM$_{00}$ mode is 1 $\mu$K before splitting and 2 $\mu$K after it. The atoms at the center of the configuration in the middle panel are due to the residual power on the TEM$_{00}$, which is required to maintain the trapping laser locked to the resonator.}
 \label{trans_c}
\end{figure}

\begin{figure}[htb]
\centering
    \includegraphics[width=0.7\textwidth]{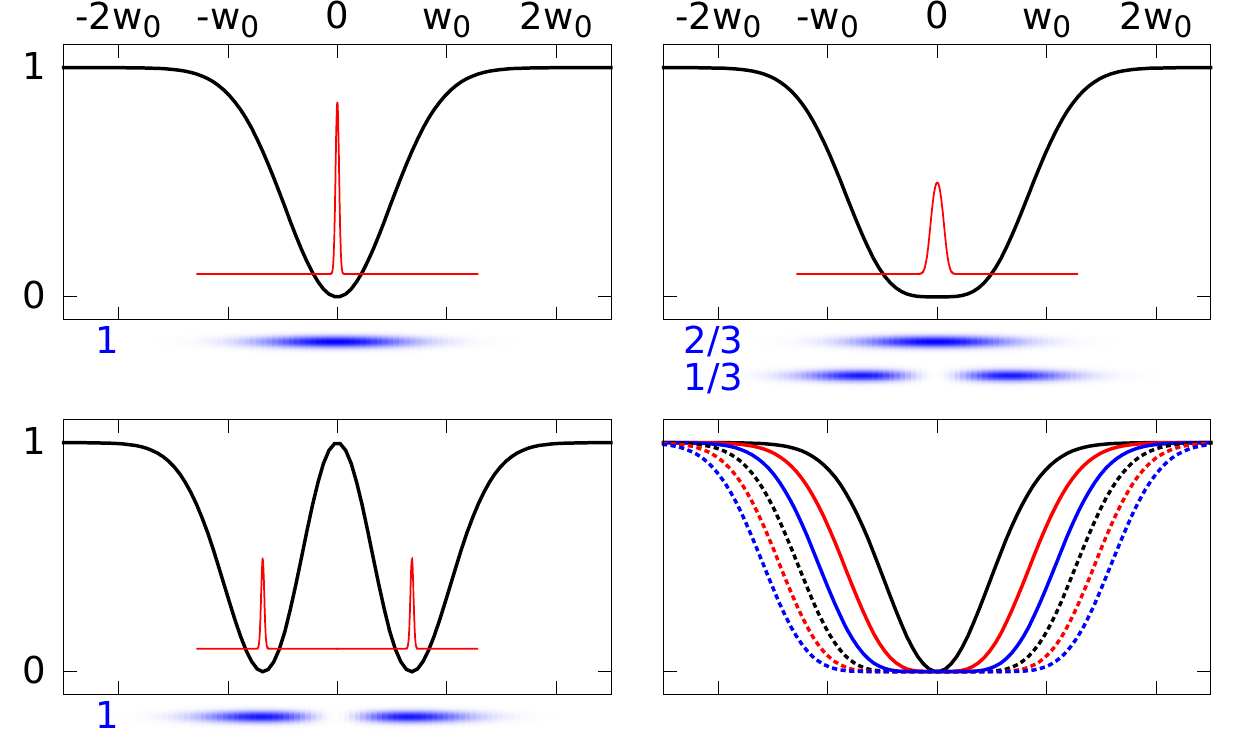}
  \caption{The beam-splitting process realized in our cavity of waist w$_0$: an initial state confined in the TEM$_{00}$ mode of the cavity (top left panel); ramping up the TEM$_{01}$ mode (see first line in Table \ref{tab:coeffTEM}, relative to quartic potential) takes the initial state to a configuration where the quadratic term in the trapping potential is zero and only the quartic term remains (top right panel), which is the point where the splitting occurs \cite{Shin_2005}; continuing the ramp process on the TEM$_{01}$ mode and extinguishing the TEM$_{00}$ leaves one with in a double well scenario (bottom left panel). An almost uniform potential in 1, 2 or 3 dimensions (see Sec. \ref{UniformPot}) can be obtained by simultaneously exciting multiple high order modes (bottom right panel). Shown are the trap potentials for one mode (TEM$_{00}$) up to six modes (TEM$_{00}$, TEM$_{01}$, ... , TEM$_{05}$) in a linear combination given by Eq. \ref{conditions} and Table \ref{tab:coeffTEM}.}
 \label{modes}
\end{figure}

\begin{table}
\caption{\label{tab:coeffTEM} Coefficients for the first 6 TEM$_{\textrm{jj}}$ modes to obtain, in the center, a 3D uniform trapping potential, whose leading order term is in the first column. Along the vertical direction the potential is homogeneous only if the linear contribution due to gravity is compensated for by the dimple.}
\begin{indented}
\item[]\begin{tabular}{@{}*{7}{l}}
\br                              
  U(x) & TEM$_{00}$ & TEM$_{11}$ & TEM$_{22}$ & TEM$_{33}$ & TEM$_{44}$ & TEM$_{55}$\cr
\mr
   x$^4$ & 1 & 1/2 & 0 & 0 & 0 & 0 \cr
   x$^6$ & 7/8 & 3/4 & 1/4 & 0 & 0 & 0 \cr
   x$^8$ & 3/4 & 13/16 & 1/2 & 1/8 & 0 & 0 \cr
   x$^{10}$ & 83/128 & 25/32 & 21/32 & 5/16 & 1/16 & 0 \cr
   x$^{12}$ & 73/128 & 183/256 & 23/32 & 31/64 & 3/16 & 1/32 \cr
\br
\end{tabular}
\end{indented}
\end{table}

\subsection{Coherent Splitting of Ultra-cold atoms} \label{beamsplitter}
Since the first experimental realization of BEC in weakly interacting atomic systems, many fundamental experiments with BEC’s have been performed. One of the more interesting prospects for BECs is their application as a source of coherent matter waves, e.g., in atom optics, atom interferometry, and quantum sensors. The application of coherent matter waves in phase sensitive experiments, like interferometers, necessitates the realization of coherent splitting of a BEC into a double-well, equivalent to an optical beam splitter. 

The most prominent matter-wave beam splitters are constructed by using either optical potentials \cite{Shin_2004,Albiez_2005,Levy_2007} or rf-dressed potentials on an atom chip \cite{Schumm_2005,Jo_2007}. Rf-dressed potentials on atom chips have shown promising results but suffer from three problems with respect to optical configurations: 1) magnetic fields can not be modulated quickly, 2) spurious effects due to switching of the magnetic fields, and 3) undesirable effects of the magnetic fields on the atomic transitions. Optical potentials bypass these problems, allowing for faster manipulation without spurious magnetic effects but are complex, requiring multiple carefully aligned beams in conjunction with advanced optics, such as spatial-light modulators and require high power to effectively confine the atoms. As our method is all-optical, it avoids the problems with rf-dressed potentials and as the cavity structure is used to create different spatial configurations, allows a simple optical arrangement with lower input power due to cavity build-up effects.

\begin{figure}[h]
\centering
\includegraphics[scale=0.50]{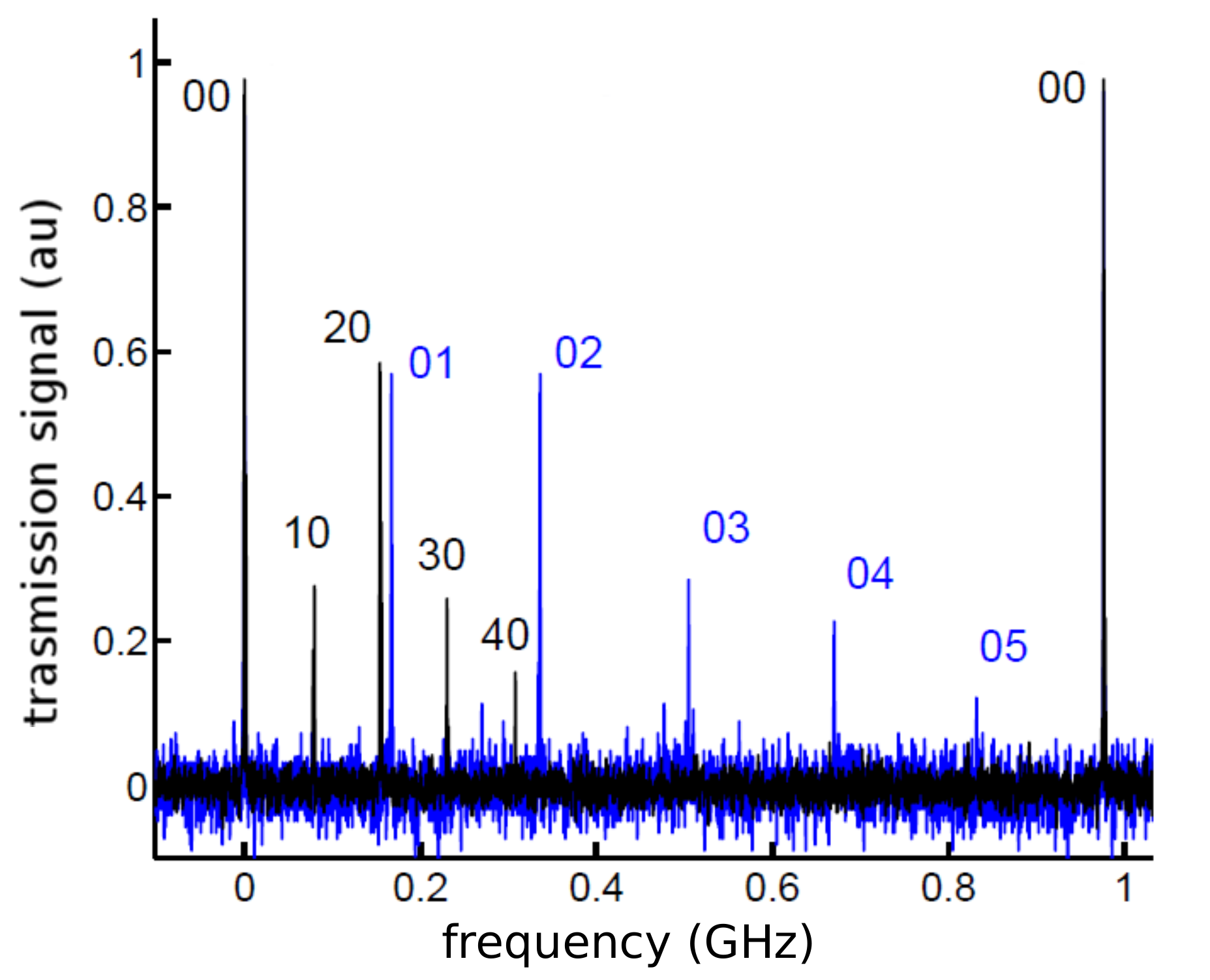}
\includegraphics[scale=0.53]{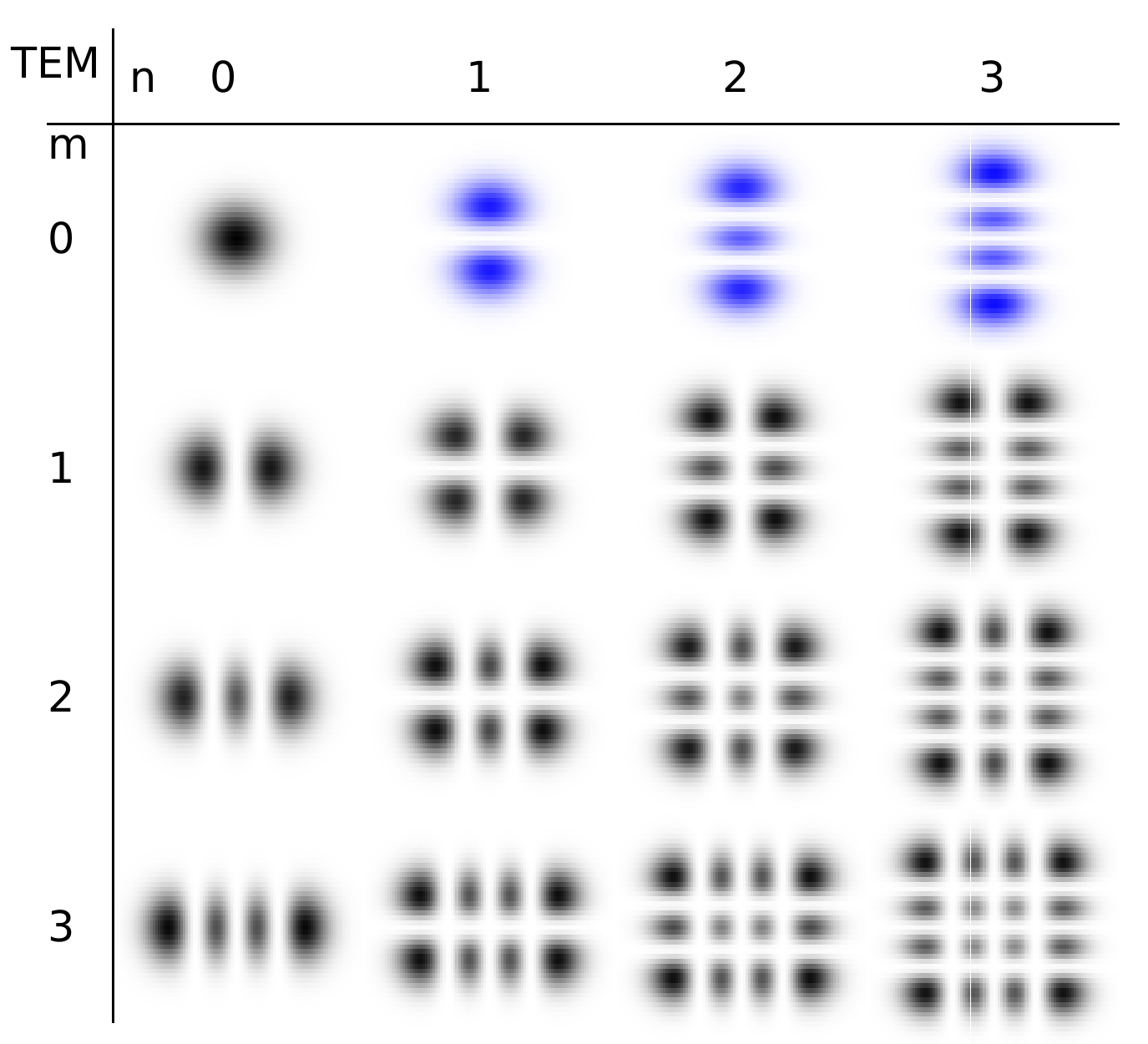}
\caption{(Top) Spectrum of the cavity transverse modes TEM$_{mn}$. The horizontal modes (black, TEM$_{m0}$) are separated by 78.9 MHz and the vertical modes (blue, TEM$_{0n}$) are separated by 164.6 MHz. All modes are non-degenerate hence selectively addressable. (Bottom) Sample modes that can be used to generate uniform potentials. For a uniform 2D potential, TEM$_{m0}$ modes (vertical column on left) can be utilized whereas for a 3D uniform potential, TEM$_{jj}$ modes (diagonal) are excited. 1D uniform potentials in the vertical direction can also be created using the TEM$_{0n}$ modes (horizontal row on top), but the linear potential due to gravity needs to be canceled, which can be accomplished by acting on the power of the dimple beam (see text). The same requirement holds for the 3D case.}
 \label{trans_nondeg}
\end{figure}

To perform the beam-splitting operation, we first create and load atoms in the TEM$_{00}$ mode as described in Sec. \ref{Exp_BECs}. After creating atoms with a temperature of 1 $\mu$K, we adiabatically excite the TEM$_{10}$ mode with an input laser beam detuned by 78.9 MHz from the beam exciting the fundamental mode and phase matched (with a phase mask) to achieve optimal coupling. At the same time we ramp down the power on the TEM$_{00}$ mode. The total power in the cavity will be given by 
\begin{eqnarray}
\bm{E}_{tot}(t) &=&  \sqrt{\alpha(t)} \bm{E}_{00} + \sqrt{\beta(t)} \bm{E}_{10} \nonumber\\
I_{tot}(t)  &\approx&  \frac{c n \epsilon_0}{2} \left[ \alpha(t) I_{00} + \beta(t) I_{10} \right] \, .
\end{eqnarray}
By adiabatically acting on $\beta$ and $\alpha$, we can coherently evolve the atoms into a double well potential -- see Fig. \ref{modes} (top left and right, bottom left). A major advantage of our cavity beam-splitter scheme is that during the entire process, odd-powered potential components are suppressed by the cavity mode structure; this is important as a necessary, but not sufficient, condition for achieving coherent beam-splitting operations. Odd-powered spatial geometries during the splitting phase mix center-of-mass and internal relative degrees of freedom, leading to mixing of internal modes, causing both decoherence and eventually heating \cite{Dobson_1994,Dalfovo_1999,BialynickiBirula_2005}. This is a potential limitation to RF and magnetic schemes \cite{Schumm_2005,Jo_2007} and to optical schemes involving beam-shaping elements \cite{Gaunt_2013}. 

We can thus increase the intensity of the TEM$_{10}$ injected light from 0 to maximum power in 60 ms, while decreasing the intensity of TEM$_{00}$, to split the atomic ensemble into four trapped ensembles. A small amount of TEM$_{00}$ must always be present as this mode provides the laser lock to the cavity for all cavity injected laser beams. Fig. \ref{trans_c} (left and middle panel) shows atoms trapped in TEM$_{00}$ and TEM$_{10}$, before and after the splitting process. We can also perform recombination by linearly decreasing the intensity of the TEM$_{10}$ beam from maximum power to 0 W and increasing that of the TEM$_{00}$ beam in 40 ms, combining the four ensembles back into the single well. The overall efficiency of the process in terms of atom transfer is greater than 95\%. The splitting and recombination process introduces a factor 2 increase in the temperature. We plan to implement an optimized ramp based on optimal control theory \cite{Hnsel_2001,Hohenester_2007,Grond_2009} to obtain the coherent splitting and recombination of a BEC using the cavity modes.

\subsection{Bose-Einstein Condensates in uniform potentials} \label{UniformPot}
New experimental tools such as Feshbach interaction resonances \cite{Chin_2010}, optical lattices \cite{Morsch_2006}, and synthetic gauge fields \cite{Dalibard_2011b}, are making ultra-cold atomic systems ideal platforms for resolving open questions in condensed matter physics \cite{Bloch_2008}. However, a non-trivial and important difference between many condensed matter systems and ultra-cold gases is that the former experience spatially uniform potentials whereas the latter are traditionally produced in harmonic traps with no translational symmetries. Nonetheless, uniform-system properties can still be extracted from a harmonically trapped sample using two methods: the local density approximation \cite{Nascimbene_2010,Hung_2011,Yefsah_2011,Ku_2012,Smith_2011}, and spatially selective probing of small portions of the ensemble \cite{Smith_2011,Drake_2012,Sagi_2012}. Unfortunately in certain situations, usually associated with diverging correlation lengths, these local approaches are not satisfying \cite{Donner_2007,Clade_2009}. In addition, uniform potentials can help in the study of equilibrium systems of strongly interacting, unitary Bose gases as a BEC in a uniform potential possess a significantly lower density than a harmonically trapped one \cite{Gaunt_2013}, reducing three-body recombination near a Feshbach resonance compared to two-body interactions \cite{Navon_2011,Li_2012}. A Bose-condensed sample in a homogeneous potential has been proposed for a new type of gravitational wave detector, based on the observation of the phonons created by a passing space-time perturbation \cite{Sabin_2014}. Therefor direct observation a spatially uniform quantum-degenerate gas would allow significant advances.

For the most part, efforts in this direction have been focused on loading atomic BECs into elongated \cite{Meyrath_2005} or toroidal \cite{Gupta_2005} traps which can be uniform along one direction, while still harmonic along the other two directions. Recently a scheme has been implemented to produce BECs in almost uniform like potentials \cite{Gaunt_2013} using a combination of three crossed blue-detuned laser beams phase imprinted via spatial light modulators. Although very impressive, this method is rather complex requiring precise positioning of the beams along with stable operation of the spatial light modulator and high optical power. In addition, creating BECs directly in a uniform potential posses problems in the evaporation stage as the densities and collisional rates are smaller, reducing evaporation efficiency. 

We propose an alternate method of creating a BEC in an almost uniform potential (1D, 2D and 3D), which can open new avenues of research in condensed matter physics. Extending the principles set forth in Sec. \ref{beamsplitter}, we propose simultaneously exciting multiple modes of our cavity, controlling the relative intensities of the individually excited modes to approach a uniform condition. By once again making use of the multiple input/output ports of our cavity, we can easily excite up to eight modes of the cavity. By creating a linear combination of the different modes, we can arrive at an almost uniform potential. This scheme offers four potential advantages to the scheme of Ref. \cite{Gaunt_2013}: 1) the structure of the cavity ensures relative alignment of the different modes; 2) the cavity build-up allows the creation of high optical potentials with low input power; 3) the TEM modes of the cavity ensures optimal spatial mode quality of trap; 4) the ease of adiabatic transfer (as shown in Sec. \ref{beamsplitter}) allows us to create the BEC in the TEM$_{00}$ mode, possessing higher trap densities thus allowing efficient evaporation, then transferring the atoms into the uniform potential. Uniform potentials in 1D, 2D and 3D can be realized depending on the modes excited.

The relative intensities of the individually excited modes can be estimated analytically. The progressive addition of each high order mode adds a spatial contour of $x^2$: so with only the TEM$_{00}$ mode, we have a maximum spatial contour $\propto x^2$, with both TEM$_{00}$ and TEM$_{01}$ we have a maximum spatial contour $\propto x^4$, and so on. The homogeneous potential is obtained by requiring that the total intensity of the $n$ cavity modes used:
\begin{equation} \label{uniform_lincomb}
I_{tot}(\bm{x}) \approx \frac{c n \epsilon_0}{2} \sum_{j=0} ^{n-1} \gamma_{jk} I_{jk}(\bm{x}) 
\begin{cases}
j = 0 & \text{1D},\\
k = 0 & \text{2D},\\
k = j & \text{3D}
\end{cases}
\end{equation}
meets the following conditions at the trap center,
\begin{eqnarray} \label{conditions}
I_{tot}(\bm{x}) \bigg\rvert_{\bm{x} = 0} &=& U_0 \nonumber\\
\frac{d^m}{dx_i^m} I_{tot} \bigg\rvert_{\bm{x} = 0} &=& 0 \hspace{0.8cm} \textnormal{$1 \leqslant m \leqslant (2n-1)$}
\end{eqnarray}
where $U_0$ is the desired trap depth, we can arrive at almost uniform like potentials with a flat trap bottom across the center and sharp boundary walls defined by the maximum spatial contour (see Table \ref{tab:coeffTEM}). The last equality is valid along all the directions $x_i$ where a homogeneous potential is sought for. In principle, the optimization only needs to be done in one of the dimensions, $x_i$, as the cavity structure ensures the same geometry in the others. Fig. \ref{modes} (bottom right) shows how the addition of high transverse modes of the cavity in the linear combination allows us to approach the uniform condition. Table \ref{tab:coeffTEM} gives the relevant terms in the linear superposition given by the constraints in Eq. \ref{conditions}.

As noted in Eq. \ref{uniform_lincomb} and in Fig. \ref{trans_nondeg}, 1D, 2D and 3D uniform potentials can be created; however, in the 1D and 3D cases (TEM$_{0j}$ and TEM$_{jj}$ modes respectively), the effect of the linear potential of gravity along the vertical direction needs to be offset, which can be easily achieved using the dimple beam (Sec. \ref{Exp_BECs}).

\section{Conclusion and future prospects}
We have demonstrated the production of multiple $^{87}$Rb BECs directly in the cavity modes. The total optical power required from the 1560 nm laser was 3 W. In the optical double well configuration we could control the BEC fraction in either of the wells by a bias magnetic field during loading process of the FORT. Also, the use of a dimple allows us to cancel the effect of gravity. Our experimental apparatus can allow us to study a wide range of phenomena which are difficult to realize in standard BEC experiments.\\

\underline{\textit{Multiple BEC cavity QED}} Some recent theoretical proposals have suggested promising schemes for the generation of entanglement of spatially separated BECs confined in a high finesse optical cavity\,\cite{Pyrkov_2013,Byrnes_2013,Joshi_2015}. Cavity assisted entanglement generation is useful for quantum information processing \cite{Byrnes_2012}. Our scheme allows us to create a BEC in a standard optical potential then adiabatically split this BEC into four BECs using the cavity resonance at 1560 nm; each of these BECs is coupled to the same optical cavity mode. This shared optical cavity mode can be used to create interactions between the BECs.\\

\underline{\textit{Atom interferometry}} This proof of principle demonstration of the splitting and the merging of atomic clouds can allow us to realize trapped atomic interferometers of a higher quality than what is currently available. We would like to use this setup to study interferometry of multiple condensates \cite{Scherer_2007,Aidelsburger_2017}. After the coherent splitting into two or more four BECs, one could implement independent engineering of the phase of each individual BEC before recombining, with the aim of studying quantum turbulence \cite{White_2014,Barenghi_2014}  and vortex formation during the recombination/interference phase \cite{Aidelsburger_2017}.\\

\underline{\textit{Strongly interacting bosons in a uniform potential}} By changing the atomic isotope from $^{87}$Rb to $^{85}$Rb, we propose to generate condensates of $^{85}$Rb in our TEM$_{00}$ cavity mode \cite{Bongs_RingCavity}. After an adiabatic transfer into the almost uniform potential and making use of the readily available Feshbach resonance \cite{Claussen_2003}, the physics of a strongly interacting, unitary Bose gas \cite{Navon_2011,Li_2012} can be studied without the need for local approximations.

\appendix

\section{Experimental set-up} \label{exp_setup}

\begin{figure}[!h]
\includegraphics[width=0.65\textwidth]{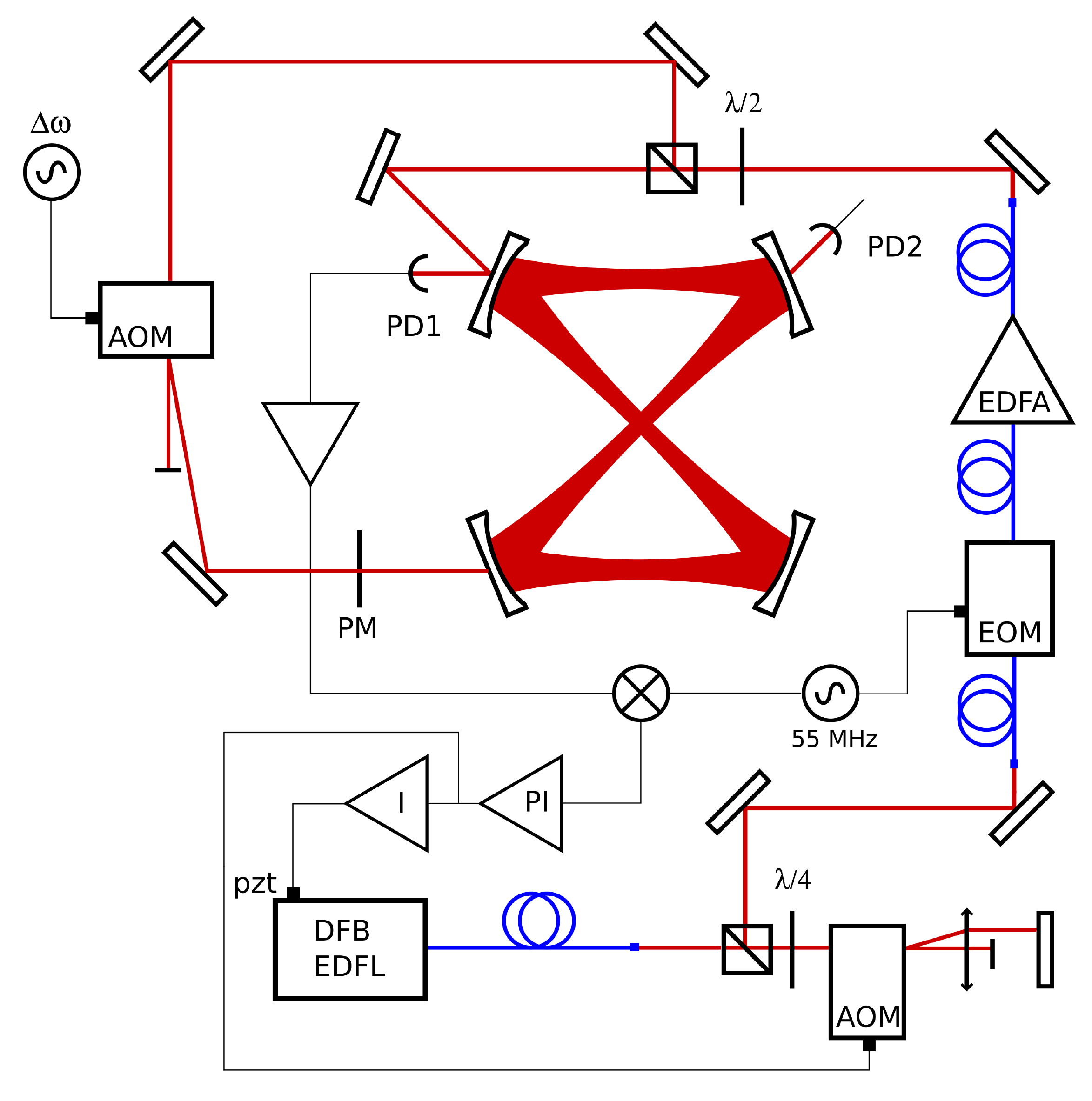}
\centering
\caption{a) A bow-tie ring cavity injected at 1560 nm is used to create the resonator enhanced FORT at the center of the cavity. The beam from the distributed feedback fiber laser (DFB EDFL) is frequency locked to the cavity by PDH locking technique. A fibered EOM is used to create 50 MHz sidebands on the beam. The relative frequency drifts are detected as beat signal on the reflection photodiode PD1. The demodulated signal is used for the feedback through a servo controller (PI). The fast and slow frequency corrections are applied to the AOM and the piezoelectric element of the laser respectively. The signal from transmission photodiode PD2 is used to reference the intensity of light in the cavity center to a stable voltage source using an intensity EOM and a low bandwidth servo controller. A vertical 1560 nm laser beam intersects the cavity center to create a dimple potential profile. The tightly focused dimple beam avoids the stagnation of the evaporation process at the end of the ramp by increasing the collision rate.}
\label{cavity}
\end{figure}

\underline{\textit{Bow-tie cavity}} The bow-tie cavity \cite{Bernon_2011} (Fig. \ref{cavity}) comprises four mirrors mounted on a titanium structure to provide mechanical stability and minimize magnetic eddy currents. Due to the square geometry (90 mm $\times$ 90 mm) the two arms of the cavity cross in the center with a relative angle of 90$^{\circ}$, which combined with the polarization being set in the cavity plane (horizontal) avoids interference patterns in the crossing region. The radius of curvature of all four plano-concave mirrors is R=100 mm yielding cavity waists for the TEM$_{00}$ mode of $w_{\parallel}$=81 $\mu$m and $w_{\bot}$=128 $\mu$m (due to the inherent astigmatism as discussed in Sec. \ref{cavitystructure}). To facilitate coarse alignment of the cavity inside the vacuum chamber, one of the mirrors is mounted on a piezoelectric actuator (Newfocus Picomotors), while another is mounted on a piezo-actuated three-axis nano-positioning system (Mad City Labs M3Z) which is used to finely adjust the cavity crossing angle and dynamically control the cavity length (it possesses a maximal angular displacement of 2 mrad and a translation of 50 $\mu$m). The free spectral range (FSR) of the cavity at 1560 nm is 976.2 MHz and the full width at half maximum (FWHM) linewidth is $\gamma$=546 kHz. The expected cavity finesse $F$=102,000 at 780 nm results in a linewidth $\sim$  4.88 kHz, opening towards the creation of correlated momentum states using the sub-recoil nature of the cavity \cite{Cola_2009,Bux_2011,Piovella_2012,Kessler2014}.

The folded geometry of our cavity invariably leads to off-axis reflections of the intra-cavity light on the mirrors. This causes the cavity astigmatism, and also gives rise to non-degenerate transverse modes well separated in frequency, e.g. $\sim$78.9 MHz between the TEM$_{00}$ and the TEM$_{01}$ modes (Sec. \ref{cavitystructure}). This can be exploited to excite multiple transverse modes simultaneously by frequency and spatial matching of different laser beams at the injection ports of the cavity. For example, a pump laser can be locked to the TEM$_{00}$ mode of the cavity (as described below) while a portion of the light can be frequency shifted by 78.9 MHz and injected into the cavity at a different input port to simultaneously excite the TEM$_{10}$ cavity mode. The relative intensity of the modes can be controlled by acting on the relative intensity of the beams at the input, whereas the coupling efficiency to each mode can be optimized by using suitable phase masks.\\

\underline{\textit{Cavity enhanced FORT}} \label{FORT} To obtain high intra-cavity power to trap neutral atoms directly from a magneto-optical trap (MOT), we use the optical resonator to enhance the 1560 nm input intensity. More precisely, with a cavity coupling efficiency of the TEM$_{00}$ mode of 35\% (15\% for the TEM$_{01}$) and an incident power of 3 W, an intra-cavity power of 250 W is obtained, giving a trapping potential with high frequencies in all spatial directions at the cavity crossing: at full power, a trap depth of U$_0$=1.4 mK is obtained with trapping frequencies of $\omega_x = \omega_y \sim 2 \pi \times 1.2$ kHz and $\omega_z \sim 2 \pi \times 1.6$ kHz when using the TEM$_{00}$ mode.

The 1560 nm light used to pump the optical cavity is derived from a single longitudinal-mode, distributed-feedback erbium-doped fiber laser (DF-EDFL) near 1560 nm (Koheras laser from NKT Photonics). The laser has a linewidth of the order of a few kHz, an output power of 100 mW single mode, linear polarization and is amplified by a 5 W erbium doped fibered amplifier (EDFA, from Keopsys). The laser is mode matched to the cavity via a beam expander, and locked to a transversal cavity mode using the PDH technique (see Fig. \ref{cavity}) where we use an AOM in a double-pass configuration (bandwidth $\sim$ 250 kHz) to realize the fast-frequency correction. Due to long term temperature variations, an additional low frequency power lock is implemented by stabilizing the light intensity transmitted by the cavity using an amplitude electro-optic modulator (EOM).\\

\underline{\textit{Light shift}} The 1560 nm trapping light is close to the rubidium 5P$_{3/2}$ - 4D$_{3/2,5/2}$  transitions at 1529 nm, resulting in a strong differential light-shift on the two levels of the D$_2$ transition. At 1560 nm, the scalar polarizability of the 5P$_{3/2}$ level is 47.7 times that of the 5S$_{1/2}$ level. As a consequence, the operation of the MOT is severely effected. In principle this effect is advantageous as the region enclosed by the 1560 nm cavity light is far detuned from the MOT cooling process and constitutes an effective `dark' region, mimicking a dark-spot MOT \cite{Clement_2009,Naik_2005}. The expected result is a higher density than normally available for $^{87}$Rb MOTs. However, as we cannot turn the 1560 nm trapping beam off without losing the lock of the laser to the cavity, the differential light-shift poses a problem for subsequent sub-Doppler cooling stages which are required for efficient loading of the laser cooled atoms from the MOT to the FORT; the `dark' regions will not participate in the sub-Doppler process.\\

\underline{\textit{Loading into FORT}} $^{87}$Rb atoms are loaded and cooled in the center of the cavity via a 2D/3D MOT setup. A two-dimensional MOT provides a source of cold atoms, with a flux of 5 $\times$ 10$^8$ at/s, directly into the cavity crossing region where we not only create our 3D MOT but we can load into our resonator FORT. During the entire MOT stage, the FORT trap depth is reduced to $\sim$110 $\mu$K. The MOT is loaded within 2 s and subsequently a compressed MOT phase (CMOT), realized by detuning the MOT beams to 6$\Gamma$ from the atomic resonance for 40 ms and decreasing the optical power by a factor 10, increases the atomic density in the FORT region. Afterwards, the cooling beams are further detuned to 43$\Gamma$ in 3 ms ramp and held for 50 ms. This mimics a sub-Doppler phase whose efficiency is limited by the strong light-shifts due to the 1560 nm beams. The high detuning was in fact chosen to red-detune the light with respect to the atomic transition in the presence of the strong light-shift. Moreover, in the FORT region, the repump light is out of resonance and the atoms are accumulated in the F=1 hyperfine state. Finally, the cooling radiation is turned off and the power in the FORT is ramped up to full power in 10 ms.

At the end of the loading sequence, we are able to load about 2$\times$10$^6$ atoms at the crossing region of the dipole trap. After a brief period for thermalization, the sample temperature is found to be 200 $\mu$K, leading to a temperature to trap depth ratio of k$_B$T/U$_0$ $\sim$ 1/6. Due to strong one-body losses lifetime of trapped atoms is limited to 6.6 s.

The measurement of the number of atoms trapped in the optical potential and of the ensemble temperature is obtained with absorption detection, taking care to previously unlock the 1560 nm from the cavity and shifting it from the cavity resonance in order to avoid its light shift effect on the D2 transition adopted for the detection. Automatic unlocking and re-locking of the trapping radiation is obtained via digital switches on the servo controller, operated synchronously with the experimental sequence.\\

\underline{\textit{Injecting Higher Order Cavity Modes}} All 1560 nm light originates from one laser. Using fiber beamsplitters, a small portion is taken and amplified to 3 W to form the TEM$_{00}$ injection beam. The laser is locked to the cavity via the Pound-Drever-Hall method \cite{Drever_1983} to this mode. The light coming from the other ports of the the fiber beam-splitters may be used to efficiently excite higher order transverse modes by 1) frequency shifting them via AOMs with respect to the TEM$_{00}$ beam by the relevant amount (Fig. \ref{trans_nondeg}), and 2) shaping their phase profile by means of a phase mask to improve the coupling efficiency. The phase mask consists of a piece of glass whose thickness varies spatially to approximate the transverse phase profile of the mode of interest. Light going through a thicker region will undergo a larger phase shift than thinner regions. For the case of the TEM$_{01}$ mode, the phase mask is simply thicker on the top half of the beam with respect to the bottom half, approximating the transverse phase distribution of that mode, whereas the phase mask of the TEM$_{10}$ mode is the same as that of the TEM$_{01}$ mode but rotated by 90$^{\circ}$.

\section{Orthogonality of the Cavity Modes} \label{app:orthogonalityModes}

The orthogonality of the $E_{mn}$ modes arises from the orthogonality condition of the Hermite-Gaussian modes:
\begin{equation}
\int_{-\infty}^{\infty} H_{m}(x) H_{n}(x) e^{-x^2} dx = \sqrt{\pi} \; 2^n \; n! \; \delta_{mn}
\end{equation}
which are orthogonal when integrated over the entire transverse $xy-$ plane of the cavity. However, atoms experiencing a linear combination of these modes inside the cavity will at any particular time only see the combined electric field of the modes at any one point within the transverse plane, experiencing an instantaneous optical potential rather than one integrated over the entire transverse plane.

At first sight Eq. \ref{eq:intprofile} does not provide an accurate description of the optical potentials; it would seem more appropriate to avoid assuming the mode orthogonality and take instead into account all of the cross terms in the calculation of the intensity

\begin{eqnarray}
\bm{E}_{tot} &=&  \sum_{m,n} \sqrt{\beta_{mn}} \bm{E}_{mn} e^{-i \; \omega_{mn} \; t}\\
I_{tot} &=&  \frac{c n \epsilon_0}{2} \lvert \bm{E}_{tot} \rvert ^2 \nonumber\\
        &=&  \frac{c n \epsilon_0}{2} \bigg\lvert \sum_{m,n} \sqrt{\beta_{mn}} \bm{E}_{mn} e^{-i \; \omega_{mn} \; t} \bigg\rvert ^2 \label{eq:fullintprofile}
\end{eqnarray}
where we added the $e^{-i \omega_{mn}  t}$ factor as our cavity is non-degenerate, implying that each particular TEM$_{mn}$ mode has a characteristic energy with a corresponding frequency, $\omega_{mn}$. Nonetheless, the condition of orthogonality can be regained by noting that as each mode has a unique frequency, the cross terms in Eq. \ref{eq:fullintprofile} will involve terms oscillating in time as $\cos(\Delta \omega \; t)$. The smallest difference in frequency between transverse modes we would like to use is $\sim$ 7 Mhz (see Fig. \ref{trans_nondeg}), meaning these cross-terms will quickly average out to zero in the relevant timescale of the atomic dynamics. Therefore Eq. \ref{eq:intprofile} would still be the valid intensity profile seen by the atoms at any point inside the cavity. It is this property which makes our cavity unique as the cavity non-degeneracy allows us to simultaneously inject many transverse modes while avoiding the non-symmetric contributions from the cross-terms of even and odd Hermite-Gaussian modes, so as to create many interesting and complex optical geometries.

\ack

We acknowledge experimental contributions from S. Bernon, T. Vanderbruggen, R. Kohlhaas and E. Cantin. This work is supported by the European Metrology Research Programme (EMRP) (JRP-EXL01 QESOCAS), Laser and Photonics in Aquitaine (APLL-CLOCK, within ANR-10-IDEX-03-02), and Horizon 2020 QuantERA ERA-NET (TAIOL).

\section*{References}
\bibliographystyle{iopart-num}
\bibliography{ms}

\end{document}